# Anomalous Hall Effect in Graphite

Y. Kopelevich[1,2] , J. C. Medina Pantoja[1] , R. R. da Silva[1] , F. Mrowka[2,*],  and P. Esquinazi[2]

[1]Instituto de Física "Gleb Wataghin", Universidade Estadual de Campinas, Unicamp 13083-970, Campinas, São Paulo, Brasil

[2]Abteilung Supraleitung und Magnetismus, Institut fur Experimentelle Physik II, Universitat Leipzig, Linnéstrasse 5, D-04103 Leipzig, Germany

ABSTRACT

We report on the experimental observation of an anomalous Hall effect (AHE) in highly oriented pyrolytic graphite samples. The overall data indicate that the AHE in graphite can be self-consistently understood within the frameworks of the magnetic-field-driven excitonic pairing models.

Experimental observations of the quantum Hall effect (QHE) and Berry phase due to Dirac-like electronic spectrum [1-4] in both multi-layered [1, 2] and single-layer [3, 4] graphite samples indicate that the physics related to the electronic properties of this material is of a broad and interdisciplinary interest. In particular, these findings suggest the interesting possibility of using graphite to test predictions of relativistic theories. One of such predictions is the magnetic catalysis (MC) phenomenon [5], which still awaits for the experimental verification. According to the quantum electrodynamics (QED) models [5, 6], an applied magnetic field B generates an electronic dynamical mass $m_{din} \sim B^{1/2}$ in both 2+1 [5] and 3+1 [6] dimensions. In the case of graphite, it is expected that a magnetic field applied perpendicular to the graphene planes opens an excitonic gap $\Delta_{El}(B)$ in the spectrum of Dirac-like fermions, associated with the electron-hole (e-h) pairing, resulting thus in a magnetic-field-induced metal-insulator transition (MIT) [7-9]. While the MIT observed in graphite [10] is consistent with theoretical predictions [7-9], additional experimental evidence supporting (or not) the MC-based scenario is necessary. In this Letter we present evidence for the anomalous Hall effect (AHE) in graphite that can be self-consistently understood within the frameworks of the excitonic pairing models which predict the occurrence of ferromagnetic correlations [7, 11, 12].

Both the Hall $\rho_H(B,T)$ and longitudinal $\rho_b(B,T)$ basal-plane resistivity measurements were performed on several highly oriented pyrolytic graphite (HOPG) samples with the van der Pauw method. The typical results obtained on two of them are reported in the present work. The samples have been thoroughly characterized elsewhere [1, 10, 13]. Briefly, the out-of-plane/basal-plane resistivity ratio at T = 300 K and B = 0, $\rho_c/\rho_b = 8.6 \times 10^3$ (sample 1, S1) and $\rho_c/\rho_b = 5 \times 10^4$ (sample 2, S2); $\rho_b = 45$ μΩ·cm (S1) and $\rho_b = 5$ μΩ·cm (S2); FWHM (full width at half maximum) = 1.4° (S1), FWHM = 0.5° (S2). In both samples Fe impurities were detected on the level of ~ 2 ppm by means of PIXE (particle induced x-ray emission) technique [13]. All dc transport measurements were made for magnetic fields applied parallel to the hexagonal c-axis in the temperature interval 100 mK ≤ T ≤ 300 K using various 9T-magnet He cryostats and a dilution refrigerator.

Figure 1 presents a selection of $\rho_H(B)$ isotherms measured for the sample S1. This figure demonstrates that at low enough temperatures $\rho_H(B)$ is a nonlinear function which can be best approximated by the equation (continuous lines)

$$\rho_H(B) = C_1\tanh(\mu B/T) + C_2 B^n, \qquad (1)$$

where $C_1$, $C_2$ and $\mu$ are fitting parameters, and $n \geq 1$. However, as the temperature increases, $\rho_H(B)$ becomes linear, the usual behavior for most normal (non-magnetic) materials.

Figure 2 provides evidence that Eq. (1) perfectly describes the data obtained for the sample S2 as well, down to T = 100 mK. Besides, Fig. 2 illustrates that the first term in Eq. (1) dominates over the second one in the low-field limit. For instance, for B = 0.03 T and T = 50 K, the ratio $C_1\tanh(\mu B/T)/C_2 B^n \approx 13$.

Interestingly, the obtained $\rho_H$ vs. B dependence at low fields agrees with the predicted functional form for the field-induced excitonic gap $\Delta_{EI} \equiv \Delta$ [7]:

$$\Delta \approx \frac{\sqrt{B}}{4\pi N}\ln\frac{\sqrt{B}}{\max[\sqrt{B}/gN;(T\sqrt{B})^{1/2}]}\tanh\frac{\Delta(B,T)}{2T} \ , \qquad (2)$$

or [9]:

$$\Delta = \frac{2T_c\sinh\dfrac{\Delta(\nu_B)}{T}}{\cosh\dfrac{\Delta(\nu_B)}{T}+\sqrt{1+\nu_B^2\sinh^2\dfrac{\Delta(\nu_B)}{T}}} . \qquad (3)$$

Here N = 2 is the number of fermion species, $g = he^2/2\pi\varepsilon v_F$ is the dimensionless parameter that characterizes the strength of the Coulomb interaction ( $\varepsilon \sim 3$ the dielectric constant and $v_F \sim 10^6$ m/s the Fermi velocity for graphite) and $\nu_B$ is the filling factor .

Equations (2) and (3) imply the existence of a critical (threshold) magnetic field $B_c(T)$ for the excitonic gap [7]:

$$\Delta \sim (B - B_c)^{1/2} , \qquad\qquad (4)$$

or [9]

$$\Delta \sim (1 - (B_c/B)^2]B^{1/2}. \qquad\qquad (5)$$

Whereas the experimental results [10] are consistent with these predictions, from the theory one obtains $B_c \approx 2.5$ T [9], a value that differs from the measured one(s) by two orders of magnitude [1, 10]. The discrepancy can be understood, however, assuming that the Coulomb interaction (g) drives the system very close to the excitonic instability. In this case, the threshold field $B_c$ can be very small or zero.

Assuming the proportionality between $\Delta$ and $\rho_H$ at low fields, we found, see Fig. 2, that $\rho_H(B)$ can be well approximated by the relations (4) or (5) below $\sim 0.2$ T, indeed. It is also instructive to verify the temperature dependence of $\rho_H$ at a fixed magnetic field. Figure 3 presents $\rho_H(T)$ data points obtained for both S1 and S2 samples in the constant applied magnetic field B = 0.03T. This figure demonstrates that the Hall resistivity $\rho_H \sim \tanh(G/T)$. Taking $G \sim \Delta/2$ , see Eq. (2), one gets $\Delta(B = 0.03T) \approx 50$ K. This implies that the excitonic gap $\Delta(B)$ is of a comparable magnitude with the distance between Landau levels of the Dirac-like electronic spectrum $\omega_L \sim 400$ [K]$(B[T])^{1/2}$ [14]. It is worth to note that the relatively large excitonic gap might be responsible for the strong damping of Shubnikov-de Haas oscillations revealed in the system of Dirac quasiparticles in graphite [2].

The relationship $\rho_H(B) \sim \Delta(B)$ can be accounted for by the theory of the field-induced excitonic state, which also predicts an extra magnetization $M \sim \Delta(B)$ in doped graphene (graphite) [7]. In this case, the Hall resistivity can be represented by the sum of two terms [15]:

$$\rho_H = \rho_{H0} + \rho_{AHE} = R_0 B + R_s \mu_0 M ,\qquad (6)$$

where the first and second terms are the Hall resistivities due to normal and anomalous Hall effects, and $R_0$ and $R_s$ are the ordinary and extraordinary (anomalous) Hall coefficients, respectively. When the second term dominates (at low fields), $\rho_H \approx \rho_{AHE} \sim M \sim \Delta(B)$, as our experimental results suggest. In the case that $\rho_H(B)$ is related to a ferromagnetic-like magnetization, we expect hysteretic behavior in certain magnetic field and temperature ranges. Indeed, measurements performed on our most disordered sample S1 revealed a small but well defined hysteresis in $\rho_H(B)$. Figure 4 illustrates this fact as well as shows the vanishing of this hysteresis at high enough temperatures. In fact, both the non-linearity and the hysteresis in $\rho_H(B)$ vanish or become immeasurably small at temperatures above approximately 150 K. We stress that the observed hysteresis in $\rho_H(B)$ invalidates classical approaches to the anomalous Hall resistance behavior [16]. At the same time, the occurrence of field-induced para- [17] or ferromagnetic-like excitonic state in graphite would naturally explain the results. Because in the H ∥ c-axis geometry graphite shows a huge diamagnetic response [18, 10, 13], a direct measurement of the paramagnetic or ferromagnetic magnetization M is not possible with conventional experimental methods. On the other hand, the magnetization hysteresis loops detected in various HOPG samples for H ∥ basal planes configuration [19, 13] suggest on the ferromagnetism in graphite with very high Curie temperature (~ 800...900 K) [19, 13]. A further work should clarify whether the AHE in graphite is related to this high temperature ferromagnetism or not.

Finally, we would like to emphasize that an occurrence of the ferromagnetism in a doped excitonic insulator state has been theoretically predicted long time ago [20], but no an unambiguous experimental confirmation of the phenomenon was found so far. The results reported in the present work strongly suggest that the applied magnetic field induces a magnetic excitonic state in graphite, although no direct experimental proof of the extra magnetization associated with the excitonic pairing can be given yet.

We thank D. V. Khveshchenko, I. A. Shovkovy, F. Guinea, and I. A. Luk'yanchuk for useful discussions. This work was supported by FAPESP, CNPq, and DFG.

Fig. 1. Hall resistivity $-\rho_H(B)$ isotherms obtained for the sample S1. Solid lines correspond to the Eq. (1) with the following parameters: $10^7 C_1 [\Omega m] = 4.95; 4.9; 4.3; 3.5; 10^7 C_2 [\Omega m T^{-n}] = 38; 31.5; 27; 22.5; \mu [K/T] = 535; 530; 560; 530; n = 1.5; 1.5; 1.7; 2.1$ for T = 15; 30; 50; 80 K, respectively; dotted line corresponds to the linear fit of the data obtained at T = 150 K; $-\rho_H(B) = aB$ with a = 9.7 $[\Omega m T^{-1}]$.

Fig. 2. Hall resistivity $-\rho_H(B)$ isotherms obtained for the sample S2 for temperatures T = 0.1; 20; 50 K. Solid lines correspond to the Eq. (1) with the following parameters: $10^7 C_1 [\Omega m] = 6; 5.5; 4.2; 10^7 C_2 [\Omega m T^{-n}] = 61; 53; 35; \mu [K/T] = 5; 700; 1000; n = 1.2; 1.1; 1.5$ for T = 0.1; 20; 50 K, respectively. Dashed-dotted and dotted lines correspond to the first and second terms in Eq. (1), respectively (T = 50 K). Dashed line corresponds to the relation (5) assuming $\rho_H(B) \sim \Delta_{EI}(B)$; the best fitting gives $B_c = 0.01$ T.

Fig. 3. Hall resistivity $-\rho_H(T)$ data points obtained for B = 0.03 T for sample S1 (●), and sample S2 (Δ) $(-\rho_H(T)/1.5)$; the line corresponds to the equation $-\rho_H(T) = A\tanh(G/T)$ where $A = 4 \cdot 10^7$ $[\Omega m]$ and G = 25 K.

Fig. 4. Hall resistivity data for sample S1; (*) $-\Delta\rho_H = \rho_H - C_1 \tanh(\mu B/T)$ at T = 5 K, $C_1 = 5.5 \cdot 10^{-7} [\Omega m]$, $\mu = 128$ [K/T]; (x) $- \rho_H(B)$ data obtained for T = 150 K after subtraction of the linear fit function, see Fig. 1. Inset shows $- \rho_H(B)$ measured at T = 5 K (*); dotted line corresponds to $- \rho_{AHE}(B) = C_1 \tanh(\mu B/T)$ for the same temperature.

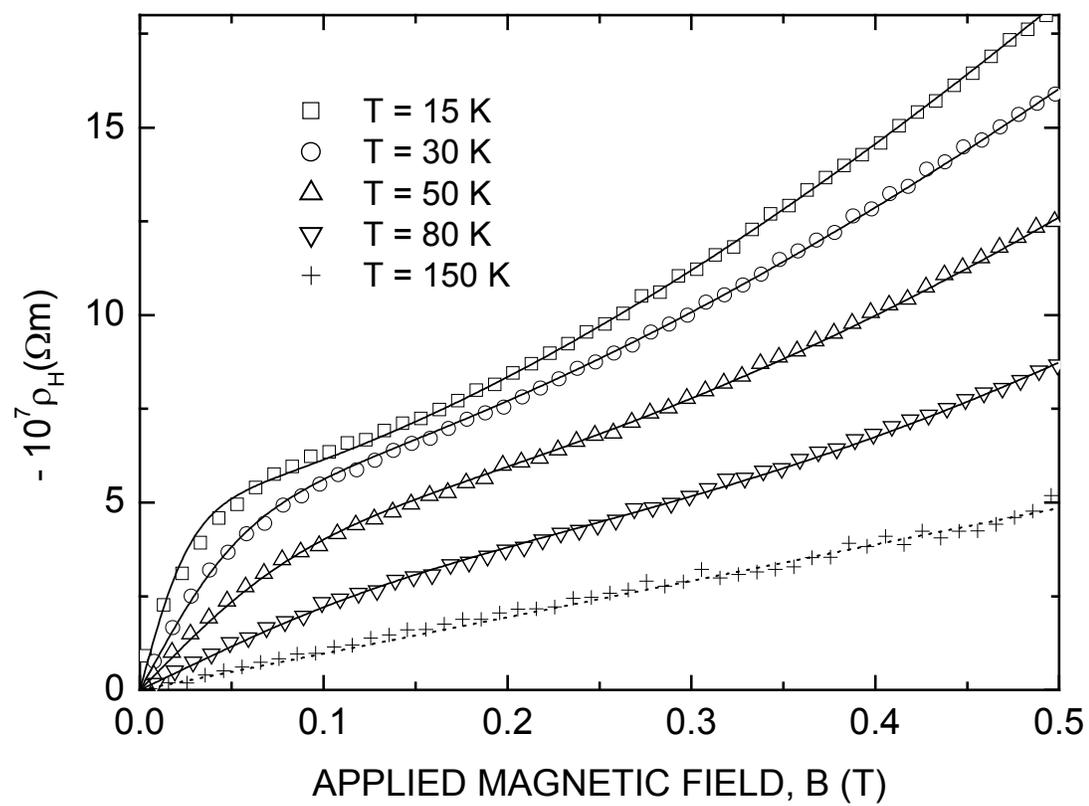

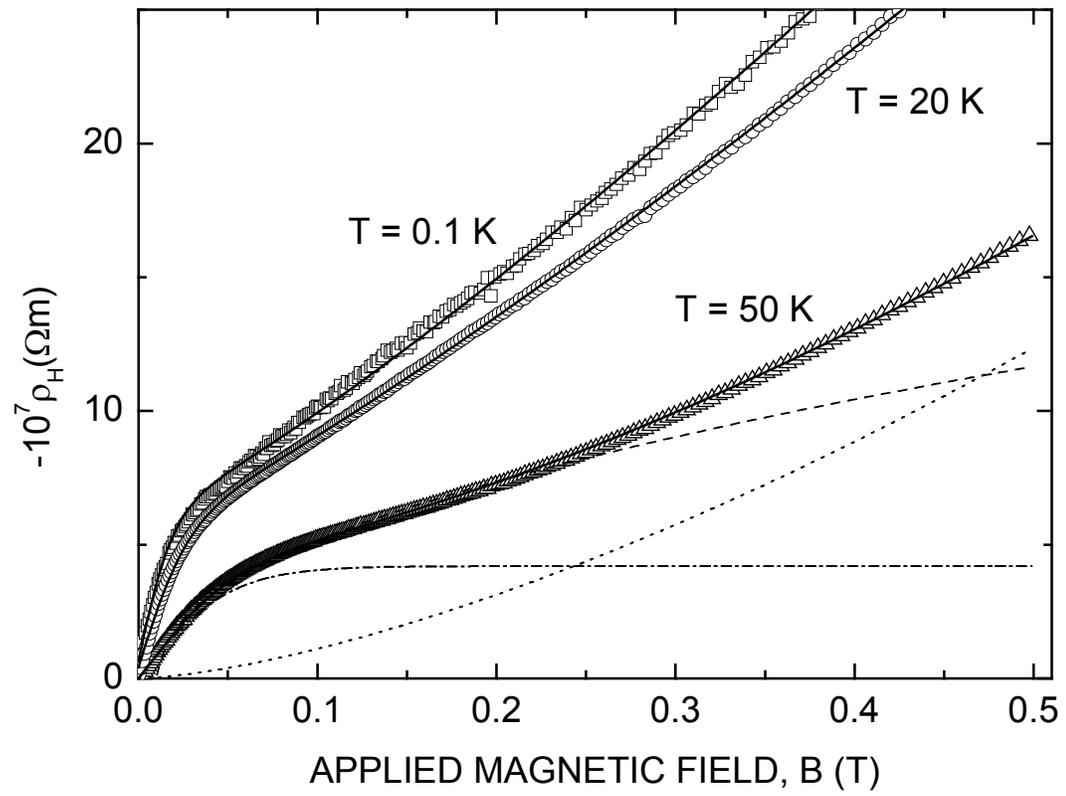

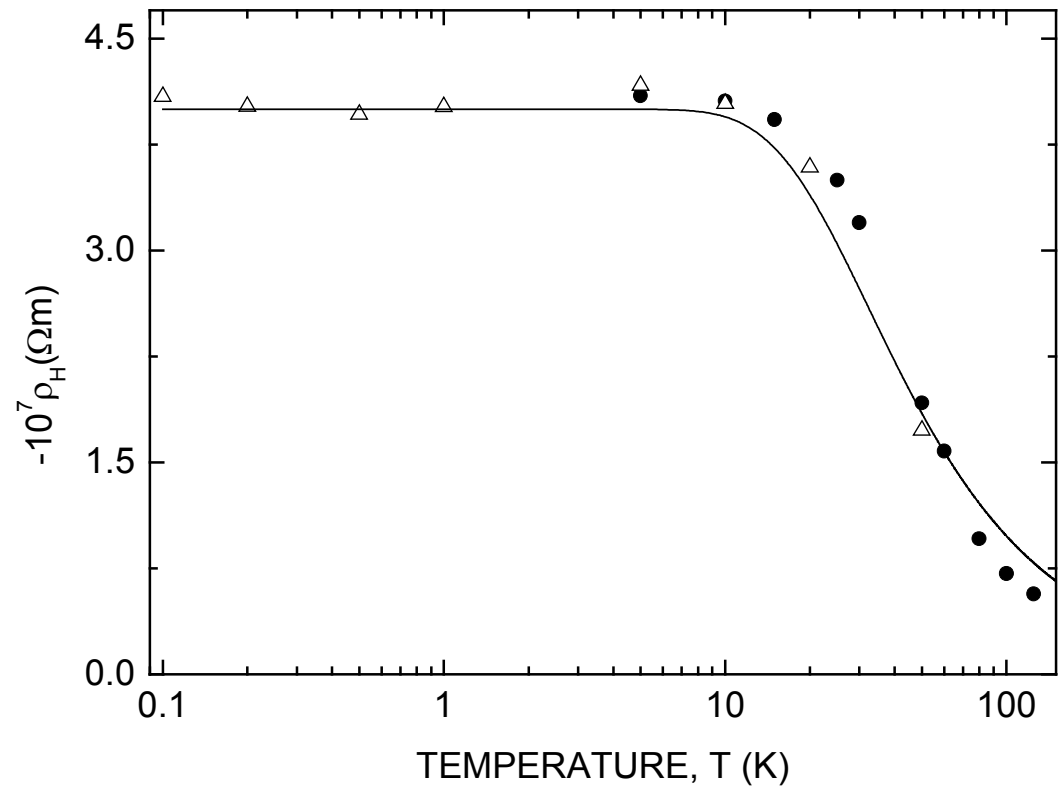

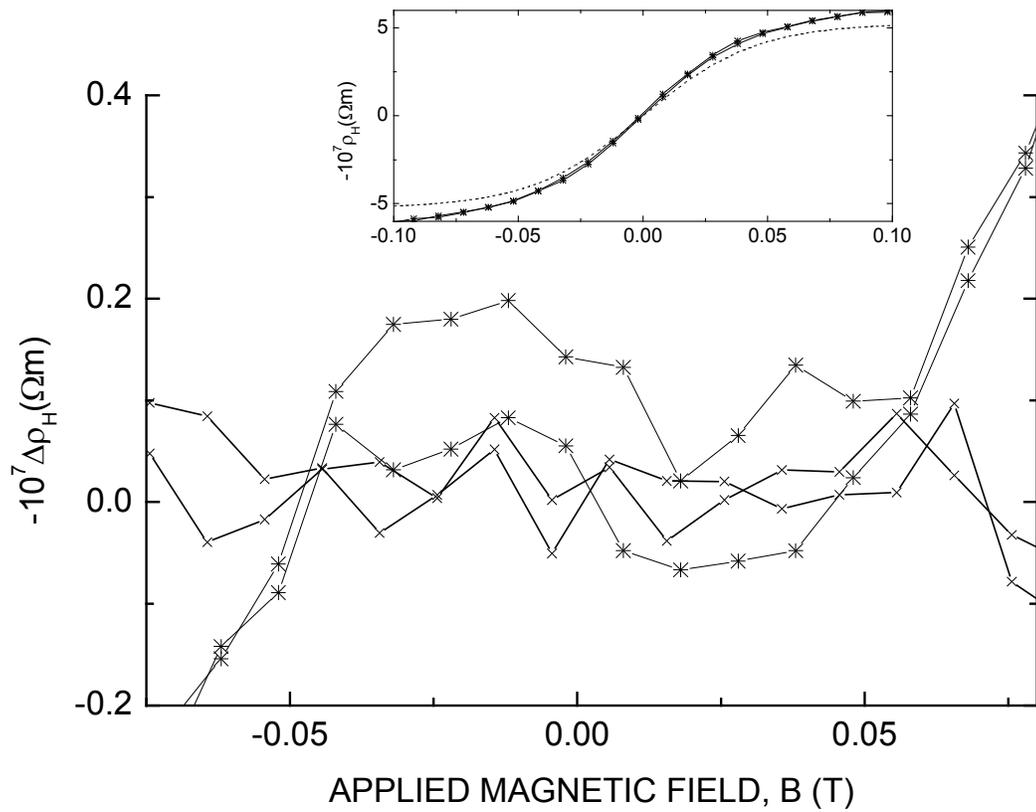

Fig. 4